\documentclass[11pt]{article} 
\usepackage{fullpage}  
\usepackage{times}  
\usepackage{url}  
\usepackage{graphicx} 
\usepackage{hyperref}
\usepackage{epstopdf}
\usepackage{soul}
\usepackage{amssymb}
\usepackage{booktabs}
\usepackage[mathscr]{eucal}
\usepackage{graphicx}
\usepackage{mathtools} 
\usepackage{tabularx}
\usepackage{microtype}
\usepackage{bm}

\DeclareMathOperator*{\argmin}{arg\,min}

\newcolumntype{Y}{>{\centering\arraybackslash}X}

\def\vd{{\bm{d}}}
\def\ve{{\bm{e}}}

\def\vg{{\bm{g}}}
\def\vh{{\bm{h}}}
\def\vi{{\bm{i}}}

\def\vr{{\bm{r}}}
\def\vs{{\bm{s}}}

\def\vv{{\bm{v}}}
\def\vw{{\bm{w}}}

\def\vy{{\bm{y}}}

\def\mA{{\bm{A}}}

\def\mD{{\bm{D}}}

\def\mL{{\bm{L}}}

\def\mP{{\bm{P}}}

\def\mY{{\bm{Y}}}
\def\mZ{{\bm{Z}}}

\DeclareMathAlphabet{\mathsfit}{\encodingdefault}{\sfdefault}{m}{sl}
\SetMathAlphabet{\mathsfit}{bold}{\encodingdefault}{\sfdefault}{bx}{n}
\newcommand{\tens}[1]{\bm{\mathcal{#1}}}

\def\tY{{\tens{Y}}}

\def\gG{{\mathcal{G}}}

\usepackage{algorithm,algorithmic}

\usepackage{color}

\title{Dynamic graph and polynomial chaos based models for contact tracing data analysis and optimal testing prescription}

\author{Shashanka Ubaru\thanks{Corresponding author, \texttt{Shashanka.ubaru@ibm.com}}\and Lior Horesh \and Guy Cohen}
\date{IBM T.J. Watson Research Center\\ Yorktown Heights, NY, USA}

\begin{document}

\maketitle

\begin{abstract}
In this study, we address three important challenges related to disease transmissions such as the COVID-19 pandemic, namely, (a) providing an early warning to likely exposed individuals, (b) identifying individuals who are asymptomatic, and (c) prescription of optimal testing when testing capacity is limited. First, we present a dynamic-graph based SEIR epidemiological model in order to describe the dynamics of the disease propagation. Our model considers a dynamic graph/network that accounts for the interactions between individuals over time, such as the ones obtained by manual or automated contact tracing, and uses a diffusion-reaction mechanism to describe the state dynamics. This dynamic graph model helps identify likely exposed/infected individuals to whom we can provide early warnings, even before they display any symptoms and/or are asymptomatic. Moreover, when  the testing  capacity is limited compared to the population size, reliable estimation of individual's health state and disease transmissibility using epidemiological models is extremely challenging. Thus, estimation of state uncertainty is paramount for both eminent risk assessment, as well as for closing the tracing-testing loop by optimal testing prescription. Therefore, we propose the use of arbitrary Polynomial Chaos Expansion, a popular technique used for uncertainty quantification, to represent the states, and quantify the uncertainties in the dynamic model. This design enables us to assign uncertainty of the state of each individual, and consequently optimize the testing as to reduce the overall uncertainty given a constrained testing budget. These tools can also be used to optimize vaccine distribution to curb the disease spread when limited vaccines are available. We present a few simulation results that illustrate the performance of the proposed framework, and estimate the impact of incomplete contact tracing data.
\end{abstract}

\section{Introduction}

Contact tracing is considered one of the most effective methods to curb the spread of {transmissible diseases such as}  COVID-19 \cite{CDCContactTracing2020}. Contact tracing is a process by which the whereabouts and interactions of an infected individual with other individuals are carefully mapped. The key information that is sought is the physical proximity between individuals and for how long the individuals interacted. Additional information such as the environment where the interaction took place (for example, a close room with poor ventilation or an outdoor space) can also be recorded.

Contact tracing can be manual or digital. Manual contact tracing is usually performed by a contact tracer, a trained health-care worker, who interviews the infected individual. Based on the infected individual's recollection of events, calendar records, credit card records, etc. the contact tracer can build a list of exposed individuals that were in proximity to the infected individual and recommend action such as quarantine or testing of the exposed individuals. 
Digital contact tracing augments the work of a contact tracer. Individuals who participate in digital contact tracing typically carry a device that tracks their proximity to other individuals. As an example, an individual’s smart phone can be used to periodically transmit a unique identifier and also record transmissions of identifiers sent by nearby devices. The signal strength of the recorded transmissions can be used to estimate the proximity to other individuals \cite{SingaporeTraceTogether2020,cohen2020crowd,DP3T2020,PEPP2020,NHS2020,alsdurf2020covi}. Digital contact tracing  that relies on this method was recently implemented by Google and Apple \cite{AppleGoogle2020} {for COVID-19} and is now available in most iOS and Android based smart phones. Digital contact tracing such as the one provided by Google and Apple is a crowd-sourcing method and its efficacy depends on adoption by the public. Furthermore, the method which uses Bluetooth transmission may inaccurately estimate of proximity due to signal attenuation or reflections from nearby objects.  However, in the workplace, on university campuses, and in schools, contact tracing can be mandated. Active or passive devices like RFID bracelets or badges that are tracked by indoor sensors, can be used inside organization's campuses to obtain reliable and accurate digital contact tracing data. For the purpose of this work we assume that contact tracing data is obtained by any of the methods discussed above.

{Some epidemics/pandemics, including  t}he recent COVID-19 pandemic have proved difficult to contain due to the large population of asymptomatic individuals. Asymptomatic people are individuals who are infected with the virus but have no symptoms. Asymptomatic people can be contagious to others. It is estimated that up to 40\% of COVID-19 infected individuals are asymptomatic \cite{li2020substantial,liu2020dynamical,weinstein2020analytic}.  Estimating the asymptomatic individuals is therefore needed to successfully curb the spread of  the disease. 
Testing  and vaccine distributions are  the other important areas that have proved to be difficult and have impeded the efforts to contain {the spread of viruses}. Without a doubt, testing is likely the most important tool that health-care professionals have to assess the spread of  viruses within the population, yet the lack of testing kits and lab resources continues to limit testing volume. Additionally the cost of testing may also limit testing in disadvantaged communities. Since testing is a limited resource, testing the entire population periodically is not feasible and therefore it is of great importance to optimally prescribe testing.      {Once effective vaccines are available, efficient vaccine distributions can curb the disease spread successfully. However, vaccine availability could be limited, and tools  to optimize vaccine distribution under limited vaccination budget are necessary.}

\paragraph{Our contributions:}
In this paper we employ contact tracing data to infer which individuals are likely to be asymptomatic and which individuals should be tested to mitigate uncertainty of the overall network. We prescribe an optimal testing recommendations to mitigate the overall risk under the constraints of limited testing resources. To achieve these goals we start by representing the contact tracing data as a dynamic graph. Each node represents an individual, and connections between the nodes represents the interaction between individuals, such as physical proximity and duration of contact. We use a compartmental epidemiological model to evolve the graph in time. The evolution also incorporates new data from contact tracing as well as new testing {and vaccination} data of individuals.

One of the epidemiological models that has been considered suitable for modeling   disease propagation is the SEIR model (Susceptible, Exposed, Infected, Recovered). This model takes into account an incubation period during which individuals that have been infected are not yet infectious themselves \cite{KermackMcKendrick1991I,KermackMcKendrick1991II,KermackMcKendrick1991III,cooper2015method,aronis2017bayesian,carroll2014visualization}. We note that our method is not tightly tied to the SEIR model and {is applicable to any other models that describe disease transmission among populations}.
The SEIR model treats the entire population as a whole and is unaware of the connections and interaction between individuals. In this work we add graphical dependency to the SEIR model equations, so the details of how individuals interact impact the model accounting for the spread of the disease. The modified SEIR model is now described using a set of partial differential equations, with a graph Laplacian operator that accounts for the interaction between individuals as captured by the contact tracing data. {Both the contact network and the paramters of the model can be time varying.} In another deviation from the original SEIR model, we treat the $S$, $E$, $I$, $R$ populations as probabilities {(similar to \cite{van2008virus,preciado2013optimal,faranda2020modelling})}, rather than compartmental populations. 

Using the aforementioned model or a similar model, it is possible to provide an early warning to individuals who are likely to be exposed or infected and also identify those individuals who are likely to be asymptomatic. The latter have a high probability of being infected while showing no symptoms. 
The second challenge that we addressed is how to prescribe optimal testing while both targeting individuals conferring eminent risk to their surrounding as well as dedicating precious testing allocation towards providing a more accurate picture of the overall risk by mitigating the overall model uncertainty. Given the large-scale nature of the problem, we propose here a Polynomial Chaos Expansion (PCE) framework to offer a rapid means for sampling the posterior distribution of the state. Quantifiable assessment of the uncertainty associated with each node in the underlying state enables identification of nodes (e.g. nodes of high variance) in the graph in which point estimate predictions can provide spurious results. It is critical to judiciously assess the degree of confidence we can attribute to our predictions, and devise means to proactively mitigate uncertainty by testing, rather than merely settle with its quantification. For this, we propose optimal testing prescription by solving an optimization problem that accounts for (a) high risk individuals according to the model, (b) the uncertainty in the model, and (c) the testing budget available.  We present simulation results that illustrate the models' behavior and show how we can issue early warnings to likely exposed/infected individuals and prescribe optimal testing to control uncertainty and mitigate the disease spread.

If effective vaccines are available, our graphical SEIR model can account for individuals who are vaccinated.  We can either incorporate another  state (say vaccinated $V$) or alternatively consolidate vaccinated with the recovered state, with slow temporal relaxation time (per the diminishing protection that a vaccine offers). We can   partition the graph and isolate communities who are vaccinated, and can ensure that the individuals who are linking between communities are vaccinated to act as buffers. We can also modify the objective of the optimization to include vaccination of individuals, taking into account the expected risk to individual and their connectivity,  the uncertainty in their state, and the amount of vaccines available, to optimize vaccine distribution. 

\paragraph{Related work:} {Since the last year}, a plethora of works have been burgeoned in the literature that model the COVID-19 disease transmission. A number of variants of the SEIR model and other transmission models have been proposed, such as the SEIR models~\cite{viguerie2020simulating} used to analysis the spread of COVID-19 in China~\cite{peng2020epidemic,tian2020global,chen2020time,liu2020dynamical,song2020epidemiological}, in Europe~\cite{faranda2020modelling,lopez2020modified,lopez2020end}, in India~\cite{biswas2020space,sardar2020assessment,gupta2020trend} and in Africa~\cite{zhao2020prediction}. Several other works exist too, that model the different aspects of COVID~\cite{berger2020seir,guerrieri2020macroeconomic,acemoglu2020multi,jones2020optimal,schwartz2020predicting,KWON2020103601,viguerie2020diffusion} and others. 
Many machine learning and AI techniques have also been explored~\cite{zhou2020artificial,mei2020artificial,vaishya2020artificial,lalmuanawma2020applications}.

Network based models have also been studied in the literature for analyzing disease spread~\cite{van2008virus,grossmann2020importance} and  optimized vaccine allocation~\cite{preciado2013optimal,bistritz2019controlling}. In these papers, a network based ODE model called the $N$-Intertwined model is proposed for analyzing the spread/transmission of COVID-19 among population. In~\cite{grossmann2020importance}, the state of the nodes is assumed to be in one of the predefined compartments, while in~\cite{van2008virus,preciado2013optimal,bistritz2019controlling} the states are stochastic. The network is assumed to be a static random graph in these models. In~\cite{preciado2013optimal,bistritz2019controlling},
 a (combinatorial) optimization problem involving a cost function of the states and constraint on the largest eigenvalue of the adjacency matrix of the graph is solved to optimize vaccination  of the population in the network.

However, to the best of our knowledge, our work is the first to incorporate contact tracing information into the SEIR model as dynamic graphs {and use the graph Laplacian for state evolution, to model the disease propagation}. This, along with with $S$, $E$, $I$, $R$ states as probabilities, enable us to issue early warnings to individuals who are likely to be exposed and/or infected (are asymptomatic). We also propose the use Polynomial Chaos Expansion to quantify uncertainties in the model and the measurements (test results) and present a method to prescribe optimal testing in order to control these uncertainties and mitigate the spread of the disease. These challenges have not been addressed in a systematic way in the prior works.

\section{Problem formulation}\label{sec:SEIR}

For the sake of simplicity we assume a population of $n$ individuals, yet,  representation of varying population size over time can also be considered.
We begin by defining {the} notation of a probabilistic individualized pandemic state tensor, its dynamics and the measurement operations.

\subsection{State}

Let the state of  individual {$i \in \mathbb{N}$} at time step $t \in \mathbb{N}$ be represented by the probability vector $\vy_{i,t} \in \mathbb{R}^4 = \{S_{i,t},E_{i,t},I_{i,t},R_{i,t} \}$, where $S_{i,t},E_{i,t},I_{i,t},R_{i,t} \in [0, 1]$ and the normalization condition applies $S_{i,t} + E_{i,t}+I_{i,t}+R_{i,t} = 1$. Thus, we assume that at each time step, an individual carries probabilities of being either susceptive, exposed, infected or recovered. The proposed framework is not restricted to the aforementioned choice, and obviously other probabilistic state representations corresponding to alternative pandemic models can equally be considered. Assuming $T$ times steps has evolved from an initial state, the state of the dynamic system is represented by the $3^{rd}$ degree tensor $\tY \in \mathbb{R}^{n \times 4 \times T}$. Incorporation of a dynamic model (even mis-specified) offers means for the incorporation of a smooth temporal prior upon the evolution of these probabilities implicitly. The state can enriched with stationary sites, such as public places, to enable transmission of disease via surface contact. Yet, proper representation of such sites may require a different state space representation as well as dedicated dynamics. 
\subsection{Measurements}

\paragraph{Graph data:}
Let $\gG_t \in \mathbb{R}^{n \times n}$ represent weighed graph data attributed to each time step. The graph represents proximity interaction between individuals. The weights on the edges factors both proximity as well as exposure duration within a single time step. Such data can be acquired from peer-to-peer short-range communication on smart devices, such as Bluetooth \cite{cohen2020crowd,bengio2020need,AppleGoogle2020,alsdurf2020covi}. Since the interactions between individuals changes over time, the set of weighted graphs forms a dynamic graph over time. We shall denote the graph Laplacian of each temporal graph $\gG_t$, by $\mL_t \in \mathbb{R}^{n \times n}$, {and is given by $\mL_t = \mD_t-\mA_t$, where $\mD_t$ is the diagonal degree matrix and $\mA_t$ is the adjacency graph obtained from the proximity/contact tracing data}.

\paragraph{Infection test data:}

In addition to the graph data, we shall assume that testing for infection  are administrated at each time step. Such tests may include PCR (Polymerase chain reaction), antibody testing such as Immunoglobulin G (IgG) or Immunoglobulin M (IgM), or any other means to assess the definitive infection state for tested individuals with measurable confidence level.
Specifically, here we are interested in tests that qualify whether an individual is actively infectious (attributed to the $3^{rd}$ components of individual's state at the timestep the test was collected). The number of such infection indicating tests (IIT) taken at each time steps may vary and given by $m_t$, whereas the results of the tests are denoted by $\vd_t \in \mathbb{R}^{m_{t}}$, $m_t<n$. For the sake of data assimilation, we denote a linear projector operator $\mP_{\vd_t} \in \mathbb{R}^{m_t \times n}$ which projects the state at time step $t$ to the IIT measurement space.

\paragraph{Recovery test data:}
Respectively, we shall denote by $p_t<n$ the number of recovery indicating tests (RIT) taken at time step $t$ and by $\vh_t \in \mathbb{R}^{p_t} $ the tests results. The RIT test qualifies whether an individual has been recovered. Similarly, as with the IIT tests, we define a linear projector $\mP_{\vh_t} \in \mathbb{R}^ {p_t \times n}$ that projects the state at timestep $t$ to the RIT measurement space.
{If and when effective vaccines are available, individuals who are vaccinated can be marked as recovered in the model.}

\paragraph{Surface test data:}
Transmission of viral content can be made via stationary surfaces, rather than merely by face-to-face interaction of individuals \cite{cirrincione2020covid, pradhan2020review}. 
It is possible to incorporate into the pandemic transmission model tracing data representing interactions between individuals and physical sites (e.g. via interaction with stationary Bluetooth device or RFID). Positive outcome of the test, will indicate that infectious particles were identified at a site. These tests can be treated similarly as IIT tests (attributed to the $3^{rd}$ components of individual's state at the time step the test was collected) or otherwise can be handled differently by augmenting the SEIR model. The number of surface tests taken at each time steps is given by $q_t$, whereas the results of the tests are denoted by $\vg_t \in \mathbb{R}^{q_{t}}$, $q_t<n$. We denote a linear projector operator $\mP_{\vg_t} \in \mathbb{R}^ {q_t \times n}$ which projects the state at time step $t$ to the surface test space.

\paragraph{Cleaning / disinfecting event data:}
When physical sites are incorporated into the model,  it is essential to indicate records of cleaning / disinfecting events which effectively reduce / reset the site to a state of having little probability of being infectious, that is annihilating the $3^{rd}$ components of individual's state at the time step the test was collected. Let the number of such recorded events be denoted by $c_t<n$ with respective recorded values $\vv_t \in \mathbb{R}^{c_t}$. The linear projector $\mP_{\vv_t} \in \mathbb{R}^{c_t \times n}$ that projects the state at time step $t$ to the disinfecting events.

\subsection{Dynamics}
To describe the dynamics of the model we modify the conventional SEIR population model, to an individualized, probabilistic graphical model. While the SEIR model has been employed extensively in disease control simulation, in the context of this study, other dynamical models can be equally utilized. Provided the interaction graph data between individuals over time as well as individuals pathogenic testing data, we shall recast the model as individualized model, where each node represents an individual, rather than address populations. Interactions between individuals and exchange of probabilities at the $t^{th}$ timestep are represented using the graph Laplacian {$\mL_t\in\mathbb{R}^{n\times n}$}. 
The revised model is a stochastic diffusion-reaction\footnote{It is important to note that other than the diffusion-reaction model considered here, alternate transport models such as wave relaxation, etc, can be considered. The discussion of such models goes beyond the scope of this study} model of the following form:
{
\begin{align}
\label{eq:cont-pde}
    \frac{d\vs}{dt} &= -\kappa_S  \mL \vs - \beta \ve \odot \vs - \gamma \vi \odot \vs + \mu_s  \vs \\
    \frac{d \ve}{dt} &= -\kappa_E \mL \ve + \beta \ve \odot \vs + \gamma \vi \odot \vs - \alpha \ve \\
    \frac{d \vi}{dt} &= -\kappa_I  \mL \vi + \alpha \ve - \mu_h \vi  -\mu_s \vs \\
    \frac{d \vr}{dt} &= \mu_h   \vi 
\end{align}
}
where {$\{\vs,\ve,\vi,\vr\}\in\mathbb{R}^n$ are vectors containing the states $\{S,E,I,R\}$ for all individuals, respectively,} $\kappa_S,\kappa_E, \kappa_I \in \mathbb{R}$ are diffusion coefficients and $\alpha, \beta, \gamma, \mu_h, \mu_s \in \mathbb{R}$ represent reaction coefficients. The model coefficients can be prescribed a-priori, but, whenever sufficient data is provided, these coefficients can be learned statistically\footnote{
Diffusion and reaction coefficients may be set a-priori differently to model individuals dynamics, vs. sites.}. The coefficients of the model themselves may evolve over time to reflect changes in individuals behaviour (e.g. masks wearing compliance, hand sanitation frequency, etc). Such refinements of the model can be accommodated by devising parametric / non-parametric models for the coefficients themselves, that includes additional health-care policies and public compliance affinity parameters. Furthermore, structural mis-specification of the dynamical model can be mitigated via hybridization of first-principle and data-driven model learning \cite{thorson2014bayesian,shulkind2018experimental}. Other then advocating for models that enables probabilistic treatment of individual state, and the incorporation of graphical data, the scope of this study focuses on closure of the tracing-sensing loop, rather than the intricacies of any particular model. Thus, for the sake of expositional simplicity we shall proceed with the above exemplar model. 

Integration of the aforementioned continuous-time dynamical system \eqref{eq:cont-pde} can be performed in various ways, such as implicit-explicit combination~{\cite{katok1997introduction}}, high order Runge-Kutta integrators~{\cite{cartwright1992dynamics}}, etc. Given the frequent rate of the graph data, and the complexity associated with semi-explicit integration schemes, we shall resort here to a simple forward Euler integration. Obviously, when such explicit integrator is employed it is essential to ensure stability of the numerical solution via careful selection of timestep duration. Other, more complex integration schemes can equally be considered. Under these settings we have: 
{
\begin{align}
    \vs_{t+1} &= \vs_t - \Delta t {\cdot}(\kappa_S  \mL_t \vs_t + \beta \ve_t \odot \vs_t + \gamma \vi_t \odot \vs_t) \\
    \ve_{t+1} &= \ve_t - \Delta t {\cdot}\left(\kappa_E \mL_t \ve_t - \beta \ve_t \odot \vs_t - \gamma I_t \odot S_t + \alpha E_t \right) \\
    \vi_{t+1} &= \vi_t - \Delta t {\cdot}(\kappa_I  \mL_t \vi_t - \alpha \ve_t + \mu  \vi_t) \\
    \vr_{t+1} &= \vr_t + \Delta t {\cdot} \mu {\cdot} \vi_t
\end{align}
}
where $\Delta t$ is the time step parameter. Note that the graph Laplacian $\mL_t$ incorporated in the model is time-varying, per the dynamic interaction between individuals over time. The contact-tracing data and the contact network are typically time varying, and it is important that the model accounts of these time dependent variations.
The initial conditions of the model are generally unknown a-priori. In the following section, we shall discuss how uncertainty associated with these conditions can be quantified and mitigated. 

{\paragraph{Asymptomatic individuals:} The proposed model assumes probabilistic state $\vy = \{S,E,I,R\}$ for  individuals, and these probabilities are estimated using the contact tracing data and the model evolution. We start with an initial state $t=t_0$, where the individuals who were tested positive will have $I_{i,t_0}=1$ and the remaining individuals start with $S_{i,t_0}=1$. Next, the model is evolved, taking into account the contact tracing data (via. the dynamic graph Laplacian) to obtain the probability states at a given time $t=T$. We can then use these probabilities to (a) issue early warning to individuals who have a high exposed state $E$ at the given time $T$, and (b) more importantly, identify  those individuals who are asymptomatic to the disease. Such individuals will have a high infected state $I$, but might not have any symptoms and hence are likely not tested. The uncertainty quantification analysis described in the following section can be further used  to identify such individuals (with uncertainty in state estimation) and prescribe testing.}
\paragraph{Data Assimilation}
    In order to provide point estimate of the state $\tY$ given measurements (testing) up till $t=T$, we can consider a dynamic inverse problem that accounts for the IIT and RIT  tests and the associated noise in the models. For example, we can consider the following problem:
    
   \begin{eqnarray} \label{eq:inverse}
    \hat{\mY}_t & =&  \argmin_{\mY =[\vs,\ve,\vi,\vr]\in \mathbb{R}^{n\times 4}} \ {\mathcal{R}}\left(\mY_{t+1}, f_t(\mY)\right)  \\
    &\textrm{s.t. } & \sum_{t=t_0}^T \eta_t {\cdot} ( \delta_d(\mP_{\vd_t}\vi, \vd_t) +    \delta_r(\mP_{\vh_t} \vr, \vh_t)\\\nonumber
    &&+\delta_g(\mP_{\vg_t} \mY, \vg_t) +\delta_v(\mP_{\vv_t} \mY, \vv_t)
    ) \le \tau
    \end{eqnarray}
    where $\eta_t \in \mathbb{R}$ represents a discount parameter, representing the degree to which one wish to factor older data (e.g. rely  more heavily on recent data rather than old one), and $\delta_d, \delta_r$ are noise models associated with  IIT and RIT tests respectively. Similarly, $\delta_g,\delta_v$ and $\mathcal{R}$ are error metrics, $f_t$ is the evolution function that takes the state $\mY_t$ to the next time step via. (5)-(8), and $\mY_{t+1}$ is the updated state after including the test results.    An alternative assimilation model would be to enforce the known testing data, rather than consolidate it with prior knowledge of disease propagation.

   Such a point estimator can be useful, yet they do not provide means for estimation of the posterior probability, and therefore, can be limiting when it comes to uncertainty quantification, and experimental design. Conventionally, one can sample the prior distribution associated with the state and update the posterior using methods such as Markov Chain Monte  Carlo (MCMC) or Hamiltonian Monte Carlo. Alternatively, methods such as generalized and arbitrary Polynomial Chaos Expansions can offer more salable means to sample the posterior in large-scale settings \cite{debusschere2004numerical,oladyshkin2012data,ahlfeld2016samba}.
    
    \section{Uncertainty Control} 
      Due to limited testing capacity, in most cases testing is performed sparsely, where the number of tests is significantly smaller compared to the dimensions of the state space $\displaystyle \sum_{t=t_0}^T m_t < \sum_{t=t_0}^T n$, rendering the state inference problem ill-posed. 
    Furthermore, the intrinsic recovery function, the interaction dynamics, and the measurements are all mis-specified, and therefore admitting uncertainty. Assuming some form of regularity of the solution (primarily in the form of the dynamical model), we can still make substantiated inferences, yet, we must consciously account for the underlying uncertainty associated with each inference. Whenever an observation (testing) takes place, one can attribute relatively high degree of confidence (small uncertainty) to the probability assigned to the relevant node, yet, the further we traverse away from that node across the graph, or propagate over time, the level of confidence decays. 
    
      Appropriate representation of uncertainty, is critical for making judicious decision as for how to prioritize best the administration of a limited testing budget. This overarching mission is essence of this study. On the one hand, it is eminent to test those identified to be under high risk (high probability of being infected), as such individuals confer immediate risk to their surrounding, yet on the other hand, acknowledging the limitation of the model, we wish to allocate testing as to reduce the degree of uncertainty associated with nodes for which uncertainty is high, as we otherwise, favor exploitation over exploration, and may miss the bigger picture altogether.

\subsection{Polynomial Chaos Expansion}
    Polynomial Chaos Expansion (PCE) is a non-sampling based formalism used for the quantification of prediction uncertainties in stochastic systems~\cite{ghanem2003stochastic,xiu2003modeling,oladyshkin2012data}.  The key idea is to depart from the traditional point-wise sampling uncertainty propagation paradigm, and instead represent the propagation of the underlying probability distribution through the stochastic process in the form of a polynomial expansion. In particular, the method
     reduces  the model into a parametric form by representing it in terms of a basis of orthonormal polynomials  with respect to the input random variables. PCE has recently been used for modeling systems in a number of applications, including machine learning~\cite{torre2019data}, sensitivity analysis of systems~\cite{crestaux2009polynomial}, flow simulations~\cite{xiu2003modeling}, geo-spatial statistics~\cite{oladyshkin2013chaos,article}, integrated circuits~\cite{kaintura2018review} and others~\cite{alexanderian2016fast,alexanderian2017mean}.  Different variants of PCE have been proposed, where the methods differ with respect to the polynomial considered~\cite{xiu2002wiener,xiu2003modeling,oladyshkin2012data}, and the approaches used for computing the coefficients~\cite{ghanem2003stochastic,debusschere2004numerical}.
    
   In this paper, we consider the arbitrary Polynomial Chaos Expansion (aPCE) approach proposed in~\cite{oladyshkin2012data}, which is a data driven approach for analyzing the stochastic (dynamical) system. The aPCE approach generalizes chaos expansion techniques to entertain arbitrary distributions with arbitrary probability measures (discrete, continuous, or discretized continuous). The expansion can be specified either analytically by virtue of probability densities or cumulative distribution functions, numerically via histograms or as discussed in the following, supported by raw data. In particular, in this study we consider the Bayesian variant of aPCE~\cite{article} were we only require knowledge of the moments of the  input random variable, rather than explicit knowledge of the probability distribution. Consider a stochastic system $\vy(\bm{\xi})$ with multi-dimensional input random variable $\bm{\xi} = \{\xi_1,\ldots,\xi_N\}$. In our case, we can consider the state  $S,E,I,R$ as four different stochastic PDE models {(as defined by eqns. (1) - (4))}, and the $N$ parameters to be the state of the $N$-nearest neighbours in the graph. Note that, the model considers the state of the neighbouring nodes to be random variables, and does not require their precise state. We wish to represent $\vy(\bm{\xi})$ by a 
   multivariate polynomial expansion as follows:
   \begin{equation}
   \vy(\bm{\xi}) \approx \sum_{i=1}^M c_i \Phi_i(\bm{\xi}),
   \end{equation}
  where the coefficients $c_i$  quantify the  dependence of the $\vy$ on the input parameters $\bm{\xi}$. The number of terms in the expansion $M$ is given as $M = \frac{(N+d)!}{N!d!}$, where $n$ is the number of parameters and $d$ is the expansion order. $\Phi_i$'s are the multi-variate orthogonal polynomial basis for $\{\xi_1,\ldots,\xi_N\}$, and assuming the parameters to be independent,  we express 
  $$
  \Phi_i(\bm{\xi}) = \prod_{j=1}^NP_j^{(\theta_j^i)}(\bm{\xi}),
  $$
  with $\sum_j \theta_j^i\leq M$ (multivariate indices that {contain the combinatorial information}). In the moment based PCE methods, the polynomials are defined as:
  $$
  P^{(k)}(\xi) = \sum_{l=1}^k \rho^{(k)}_l \xi^l, \quad l\in [0,d],
  $$
  where $\rho^{(k)}_l$ are {the} coefficients of $P^{(k)}$ {for a variable $\xi$}. 
    
    The method in~\cite{oladyshkin2012data,article} constructs these polynomials for any arbitrary distributions by just using the moments computed from observed/sampled data. Suppose we have $T_0$ observations of the data $(\vy,\bm{\xi})$, we can compute the raw moments $\mu_l  = {\tfrac{1}{T_0}\sum_{t=1}^{T_0}\xi_t^l}$ for $l=0,\ldots, 2k-1$. Then, the coefficients $\rho^{(k)}_l$ of the polynomials $P^{(k)}$ are computed by solving a linear system with the following square matrix of moments, see~\cite{oladyshkin2012data} for details.
    \begin{equation}
        \begin{bmatrix}
            \mu_0 & \mu_1& \cdots & \mu_k\\
            \mu_1 & \mu_2& \cdots & \mu_{k+1}\\
             \vdots & \vdots& \vdots& \vdots\\
                   \mu_{k-1} & \mu_k& \cdots & \mu_{2k-1}\\
                    0 & 0& \cdots &1\\
        \end{bmatrix}
             \begin{bmatrix}
            \rho^{(k)}_0 \\ \rho^{(k)}_1 \\ \vdots \\ \rho^{(k)}_{k-1}\\ \rho^{(k)}_k 
        \end{bmatrix} 
        =
       \begin{bmatrix}
            0 \\ 0 \\ \vdots \\ 0\\ 1
        \end{bmatrix} 
    \end{equation}
    Once the polynomials $P^{(k)}$ are constructed from the moments of sampled inputs $\bm{\xi}$, the   coefficients $c_i$ can be computed for the observed data $\vy$ using Gram–Schmidt orthogonalization  or by the Stieltjes procedure (solving a least squares problem). 
    The coefficients can be then updated using a Bayesian approach for the additional observed/sample data, see~\cite{article}.

    In our case, PCE treats the state $\{S,E,I,R\}$ evolutions as stochastic dynamic systems, and tries to model the probability distribution of the states. The measurements correspond to the testing results $\vd_t$ and $\vh_t$. Once, we obtain the PCE, we can compute the posterior statistics such as the posterior mean $\hat{\mu}$ and variance $\hat{\sigma}$ for the output model, inexpensively {by} simply constructing the response surface using the coefficients of the polynomial expansion. In our case, we can obtain the posterior mean and variance for the four states for each individual using aPCE. The posterior statistics can then be used to identify uncertainties in the individual's states, and optimal testing  can be prescribed.

    \subsection{Optimal testing prescription}
    One of the main challenges related to pandemics has been the issue of prescribing testing optimally given limited testing resources. The aPCE approach described above helps us quantify uncertainty, and  using the posterior statistics, we can prescribe optimal testing to control/mitigate the uncertainty.
    
    Suppose the probability associated with each state $\vy_t$  be denoted by a $2^{nd}$ moment construct accounting for both the mean probability $\mu_t$ and the variance $\sigma_t$, representing the state uncertainty, i.e. $\vy_t \sim \mathcal{N} \{\mu_t, \sigma_t I \}$.  
    Our goal would be to figure out what is the best testing paradigm in the next time step, so as to (a) minimize the risk of infection propagation, while also (b) minimize the uncertainty associated with the state, and (c) account for the limited testing budget. Let, $\vw_t \in \mathbb{R}_+^n$ denote the recommended testing assignment for the time step $t$. Then, we propose to solve the following test allocation problem:
    \begin{eqnarray}\label{eq:opt} 
    &\hat{\vw}_t &= \argmin_{\vw_t} \left\{ U(\vw_t,  \hat{\sigma_t}) \ +  D(\vw_t,{A}(\hat{\mu_t},\hat{\sigma_t}),\vd_t) +\lambda \|\vw_t\|_1 \right\} \\
    & \textrm {s.t. } &  0\leq w_{i,t}\leq 1 \quad i\in [N]
    \end{eqnarray}
    where function $U(\cdot)$ represents the posterior uncertainty (measured using the posterior of variance $\hat{\sigma}_t$ computed using PCE) associated with performing tests per $\vd_t$, and function $D(\cdot)$ captures the degree in which testing should be performed to those who are in the highest risk of being infected (a form of bias-variance balance), with $A(\cdot)$ is an acquisition function that quantify the discrepancy between infected symptomatic and asymptomatic individual. The posterior mean $\hat{\mu}$ and variance $\hat{\sigma}$ are computed using the PCE estimate. The $\ell_1$ regularization is used to control the sparsity of $\vw_t$, i..e, the number of tests to be performed at time $t$, based on the testing budget available. $\ell_0$ (quasi) norm cardinality constraint  can also be used for a bounded test budget, say $\|\vw_t\|_0\leq k_t$, where $k_t$ is the maximum number of tests available at time $t$ \cite{van2009probing}.
    We can also split the problem into two separate minimization problems in order to assign predefined budget to the two criteria (risk and uncertainty). In our simulation experiments,  Euclidean norm error function was used for  posterior uncertainty $U(\vw_t,  \hat{\sigma_t}) = \|\vw_t - \hat{\sigma_t}\|_2$,  the upper confidence bound $A(\hat{\mu_t},\hat{\sigma_t}) = |\hat{\mu_t}- \beta\hat{\sigma_t}|$ was used as the acquisition function, and the distance measure was $D(\vw_t,a(\hat{\mu_t},\hat{\sigma_t}),\vd_t)  = \| \vw_t -  a(\hat{\mu_t},\hat{\sigma_t}) \odot (1 - \vd_t)\|_2$.

    We also wish to remark here that, when effective vaccines are available for distribution, we can modify the above optimization problem to obtain \emph{optimal vaccine allocation}, in order to curb the disease spread and account for the amount of vaccines available at a given time.

   {\paragraph{Detailed Algorithm:} Here, we present the detailed procedure for the proposed model. The overall algorithm has three main stages. In the first stage, for each time $t$, the contact tracing data (dynamic graph $\mL_t$) and current testing results ($\vd_t,\vh_t$) are used to evolve the state $\mY_t = [\vs_t,\ve_t,\vi_t,\vr_t]$ using the dynamic graph SEIR model and eqns (5)-(8). In the second stage, using the set of $T$ observations $\tY$, we build polynomial chaos expansions for the four state $S,E,I,R$, and the PCE estimate 
   $\hat{\tY} \in\mathbb{R}^{n\times4\times T}$ (response surface) is computed. Finally, in the third stage, using the estimate  $\hat{\tY}$, we compute the posterior mean and variance $\hat{\mu}_T,\hat{\sigma}_T$, and solve the optimization problem in eqn.~\eqref{eq:opt} to obtain the optimal testing prescription $\vw_T$. 
    \begin{algorithm}[H]
\caption{Dynamic graph and polynomial chaos based models for disease propagation and optimal testing  }
\label{alg:algo1}
\begin{algorithmic}
{
   \STATE {\bfseries Input:}  Test results $\vd_t,\vh_t$, Graph Laplacians $\mL_t$ for time $t=0,\ldots,T$, model parameters, $N, d$
 \STATE {\bfseries Output:} Optimal testing vector $\vw_T$.
  \STATE  {\bfseries  0.}  $\mY_0 = $ initializeStateSEIR( $\vd_0,\vh_0$).
  \STATE  {\bfseries Stage I}
 \FOR{$t=1,\ldots, T$}
 \STATE  {\bfseries  1.} $\mY_t =$ updateStateTestingSEIR($\mY_{t-1},\vd_t,\vh_t$).
  \STATE  {\bfseries  2.} $\mY_t =$ evolveGraphSEIRModel($\mY_{t},\mL_t$, parameters)
  \STATE  {\bfseries  3.} Issue early warnings to exposed individuals.
\ENDFOR
  \STATE  {\bfseries Stage II}
 \FOR{$i=1,\ldots, 4$}
\STATE {\bfseries  4.} $[\bm{\Xi },\mZ] = $ NeighborState($\mL_T,\tY(:,i,:),N$).
\STATE {\bfseries  5.} $\hat{\tY}(:,i,:) =$ Bayesian-aPCE($\bm{\Xi },\mZ,\vd_T,\vh_T,N,d$).
\STATE {\bfseries  5.} If $i=3$,  estimate asymptomatic individuals.
\ENDFOR
  \STATE  {\bfseries Stage III}
\STATE {\bfseries  7.} $[\hat{\mu}_T,\hat{\sigma}_T] = $
weightedMeanVariance($\hat{\tY}$).
\STATE {\bfseries  8.} $\vw_T =$optimalTesting($\hat{\mu}_T,\hat{\sigma}_T,\vd_T$)

}
\end{algorithmic}
\end{algorithm}
   
  Algorithm~\ref{alg:algo1} describes the procedure. The function `initializeStateSEIR' initialize the state based on the initial test results $\vd_0,\vh_0$ (i.e., set $Y_0(i,3)=1$ if $\vd_0(i)= 1$; $Y_0(i,4)=1$ if $\vh_0(i)= 1$; else $Y_0(i,1)=1$), and `updateStateTestingSEIR' function updates the state $\mY_t$ based on the test results $\vd_t,\vh_t$  (i.e., set $Y_t(i,3)=1$ if $\vd_t(i)= 1$; and $Y_t(i,4)=1$ if $\vh_t(i)= 1$).
   Next, the function `NeighborState' find the states of the $N$ neighbors ($N$ input variables) for each individual $\bm{\Xi}\in\mathbb{R}^{n\times N}$ given the current graph Laplacian $\mL_T$, and the output samples $\mZ\in\mathbb{R}^{n\times T}$. Then, the `Bayesian-aPCE' function constructs arbitrary polynomial chaos expansion and outputs an estimate (response surface) for $\mZ$. The `weightedMeanVariance' function computes the mean and variance across time of the response surfaces of the four states, and then computes a weighted mean of the four states, where states $I$ and $E$ are weighted more than the other two states, since we are more interested in the uncertainty of these states.
  } 

\section{Simulation Results}
In this section, we present few numerical results based on simulations\footnote{Much of real-world contact tracing data are private and are not publicly available. Our simulation results show how our methods can be deployed on contact tracing data.} to illustrate the behaviour of the different aspects of our models. We first show how the graphical SEIR model captures the disease dynamics, and how we can use it to issue early warnings to individuals who are likely infected/exposed. We then show how aPCE and uncertainty quantification can be used to prescribe optimal testing, when the testing resources are limited. 

     \begin{figure*}[tb!]
\centering
\includegraphics[height=0.345\textwidth,trim={1.0cm 1.1cm 1.0cm 0.55cm}]{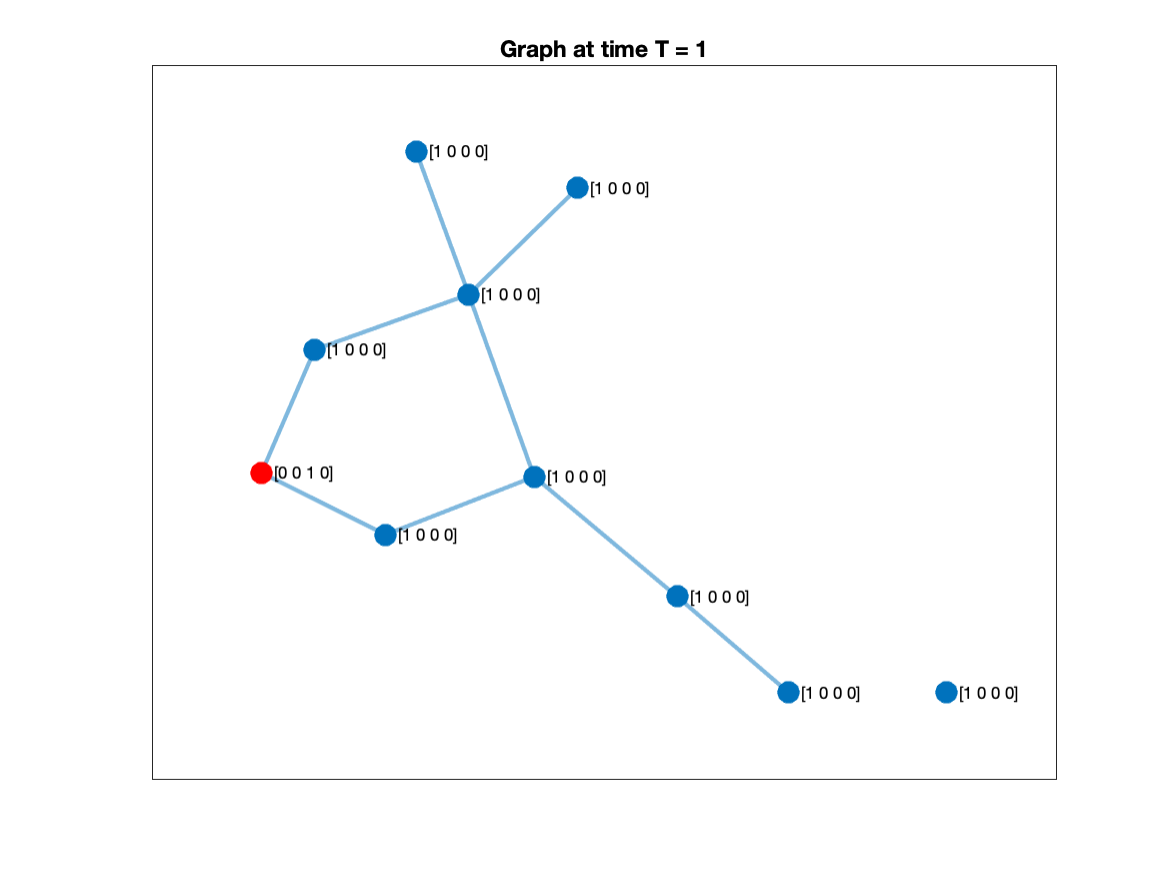}
\includegraphics[height=0.345\textwidth,trim={1.0cm 1.1cm 1.1cm 0.55cm}]{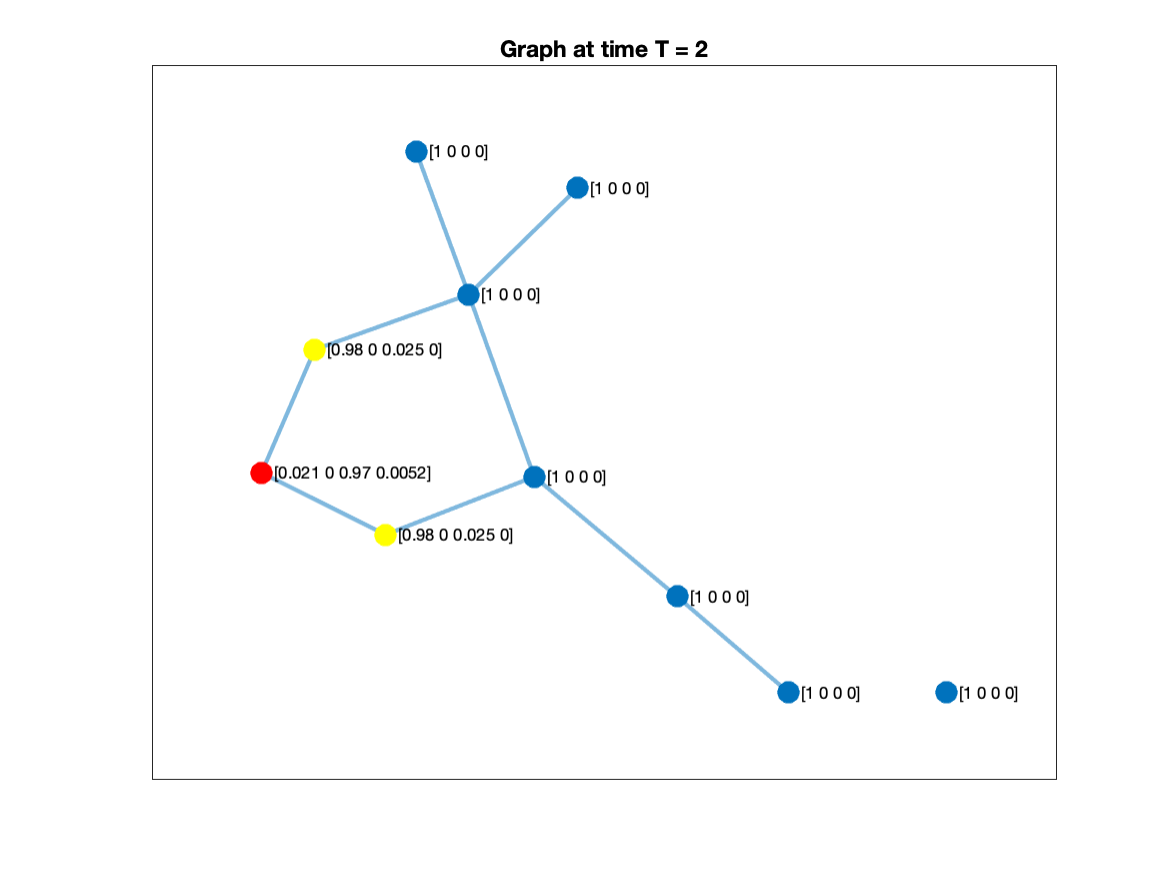}
\includegraphics[height=0.345\textwidth,trim={1.0cm 1.1cm 1.0cm 0.55cm}]{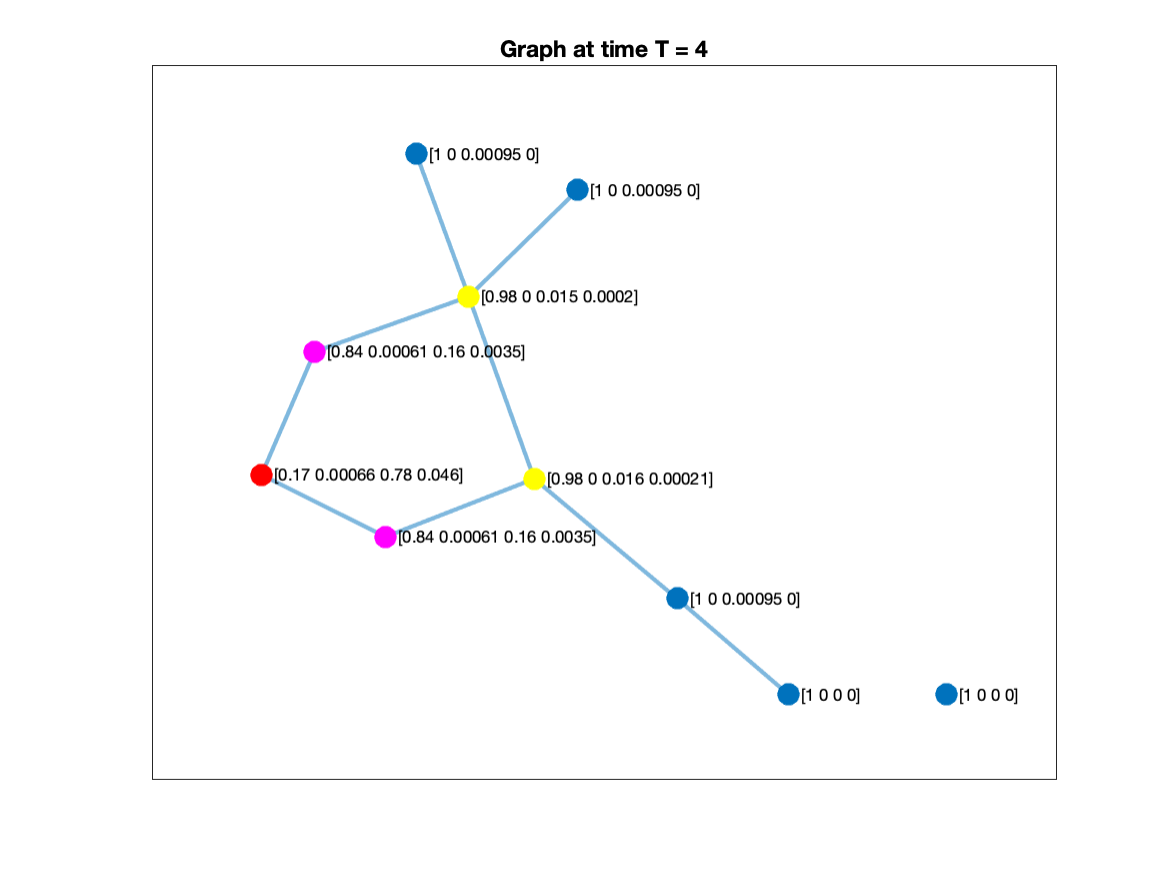}
\includegraphics[height=0.345\textwidth,trim={1.0cm 1.1cm 1.1cm 0.55cm}]{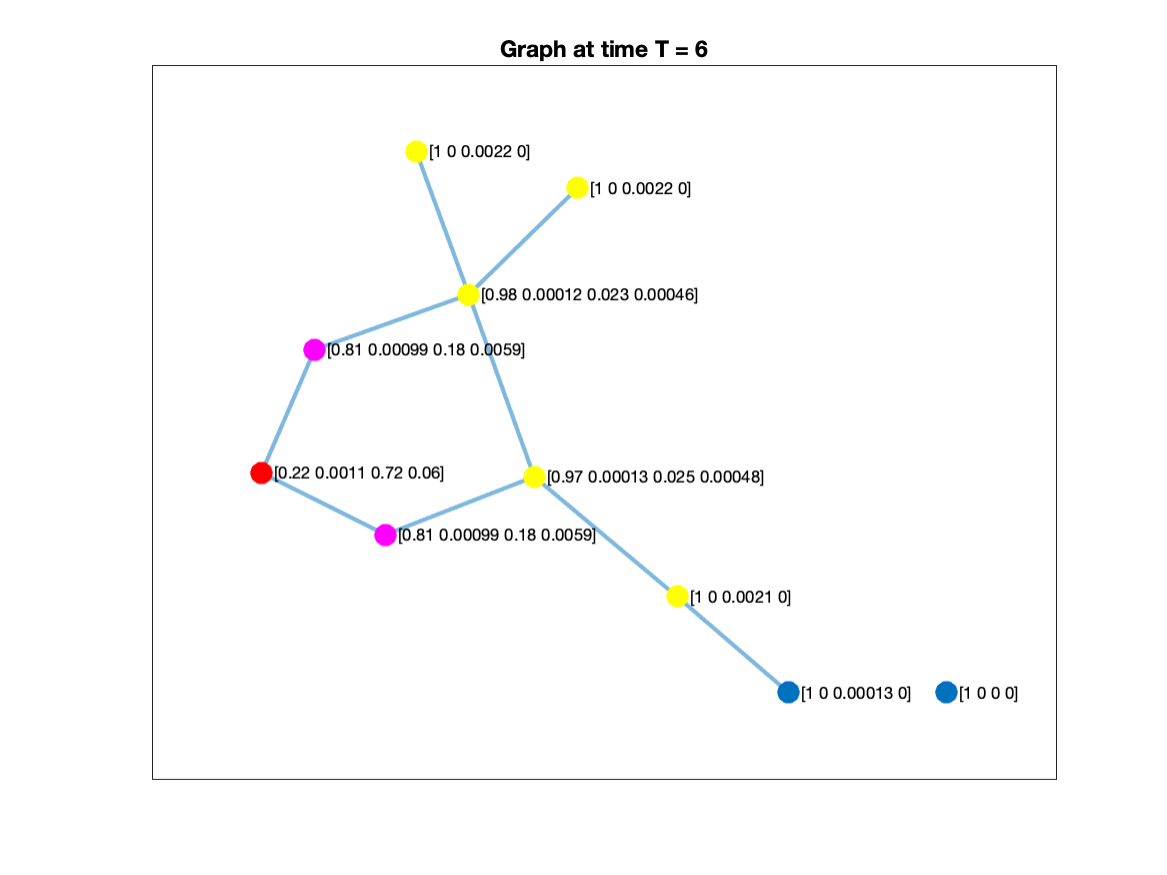}
\includegraphics[height=0.345\textwidth,trim={1.0cm 1.1cm 1.0cm 0.55cm}]{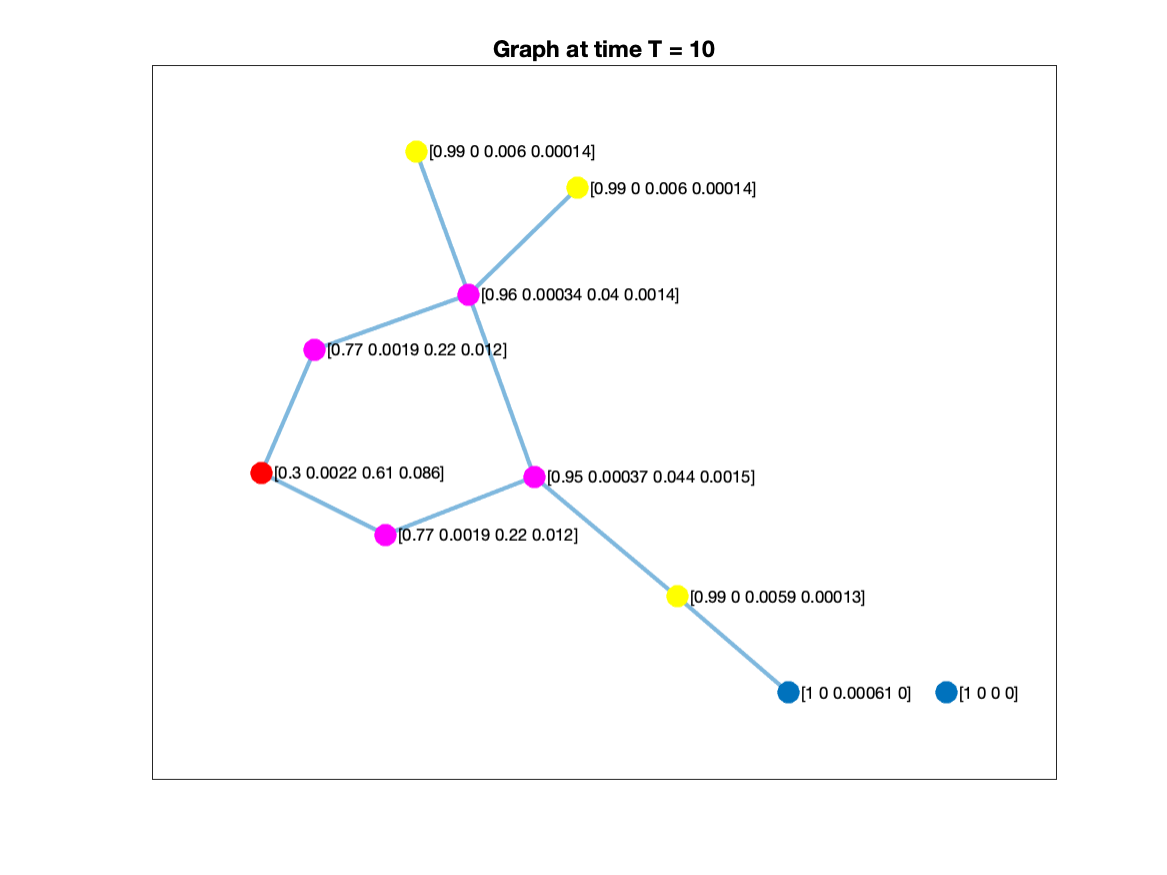}
\includegraphics[height=0.345\textwidth,trim={1.0cm 1.1cm 1.1cm 0.55cm}]{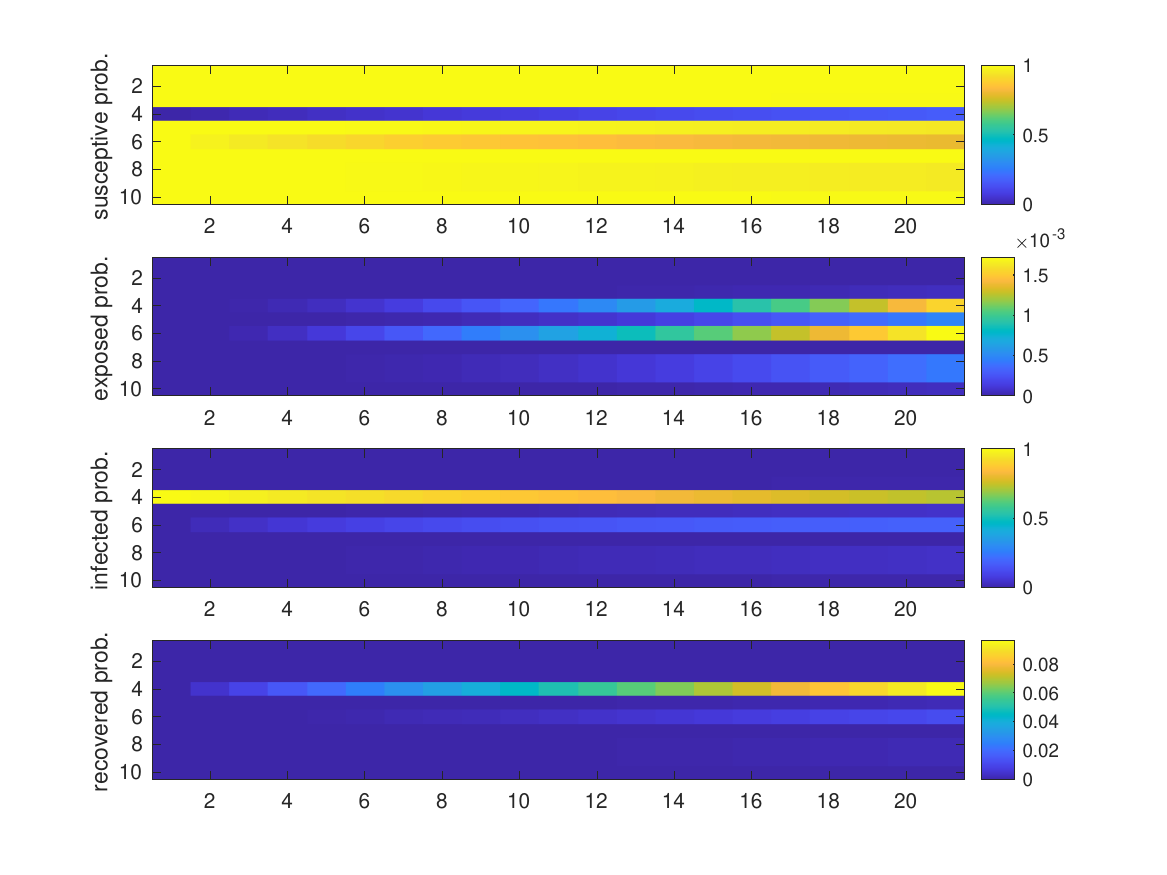}
\caption{Graphical SEIR model disease transmission visualization. Sample simulation with 10 nodes at five time instances  (first five images). Red nodes indicate infected individuals $I >0.5$, magenta nodes have $I>0.04$, and yellow have $I>0.002$. The last plot depicts the state $\{S,E,I,R\}$ for the 10 individuals  over 20 time instances.} 
\label{fig:graphs}
\end{figure*}

\paragraph {Graphical SEIR model:} In the first set of experiments, we analyze the graphical SEIR model proposed in section~\ref{sec:SEIR}. In figure~\ref{fig:graphs}, we illustrate the disease transmission as modelled by the graphical SEIR model. We consider a small (fixed) graph of 10 individuals (for easy visualization) and show how the infection transmits to other nodes over time. At time step $t=1$, we have one individual infected (red node). We note that as time evolves, the infection spreads to nodes who are at close proximity. We consider a fixed graph here for illustration, but a graph that varies over time (better simulation of human interactions) is considered in the remaining experiments.  We note that the state of the nodes evolve over time as the virus spreads. As examples, we have magenta nodes with $I>0.04$, and the yellow nodes with $I>0.002$, and we note the change of states over time. Based on this model, we can \emph{issue early warnings} to the individuals (via. text messages or app notifications) if their state $I$ or $E$ crosses certain thresholds, possibly even before the individuals show any symptoms. In our example,
we can send out warnings to the individuals, when their colors change, once when blue to yellow and again when yellow to magenta. 

\begin{figure*}[tb!]
\centering
\includegraphics[height=0.345\textwidth,trim={1cm 1cm 3cm 0.5cm}]{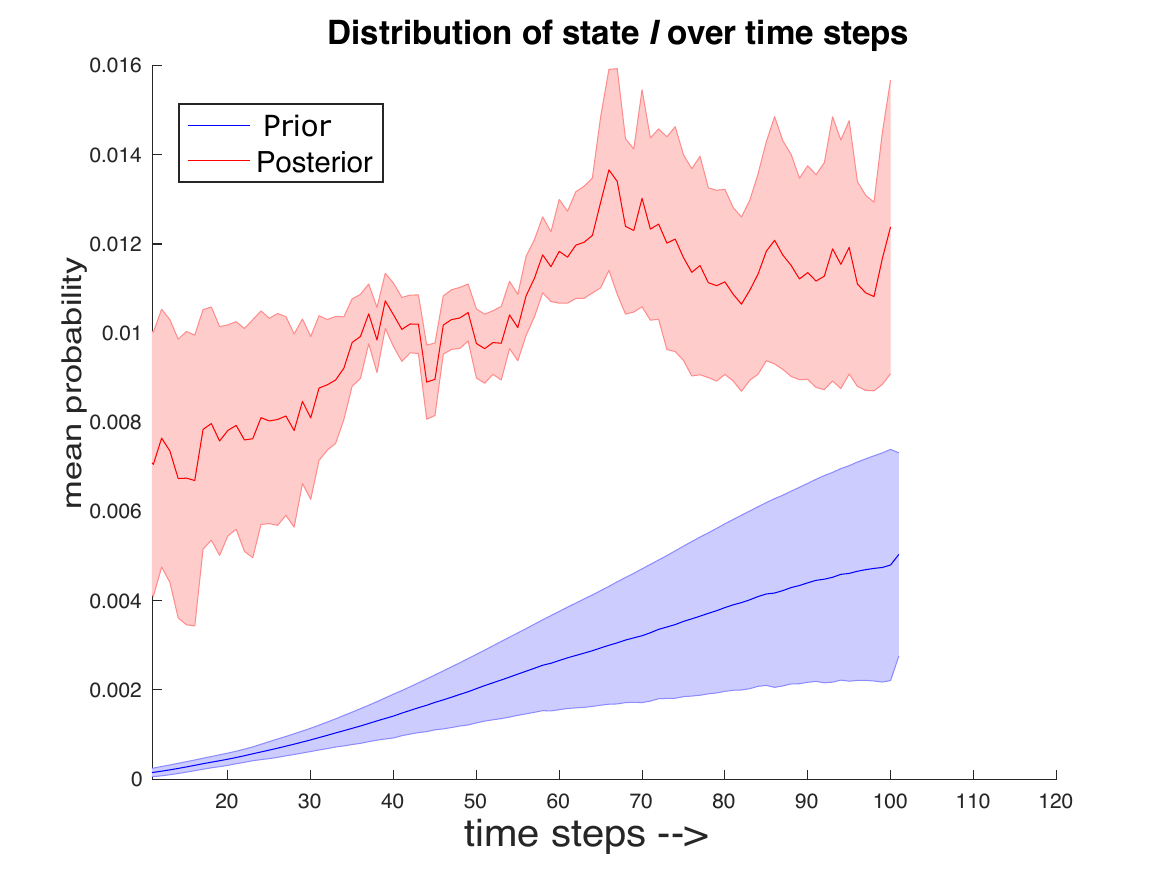}
\includegraphics[height=0.345\textwidth,trim={1cm 1cm 0.5cm 0.5cm}]{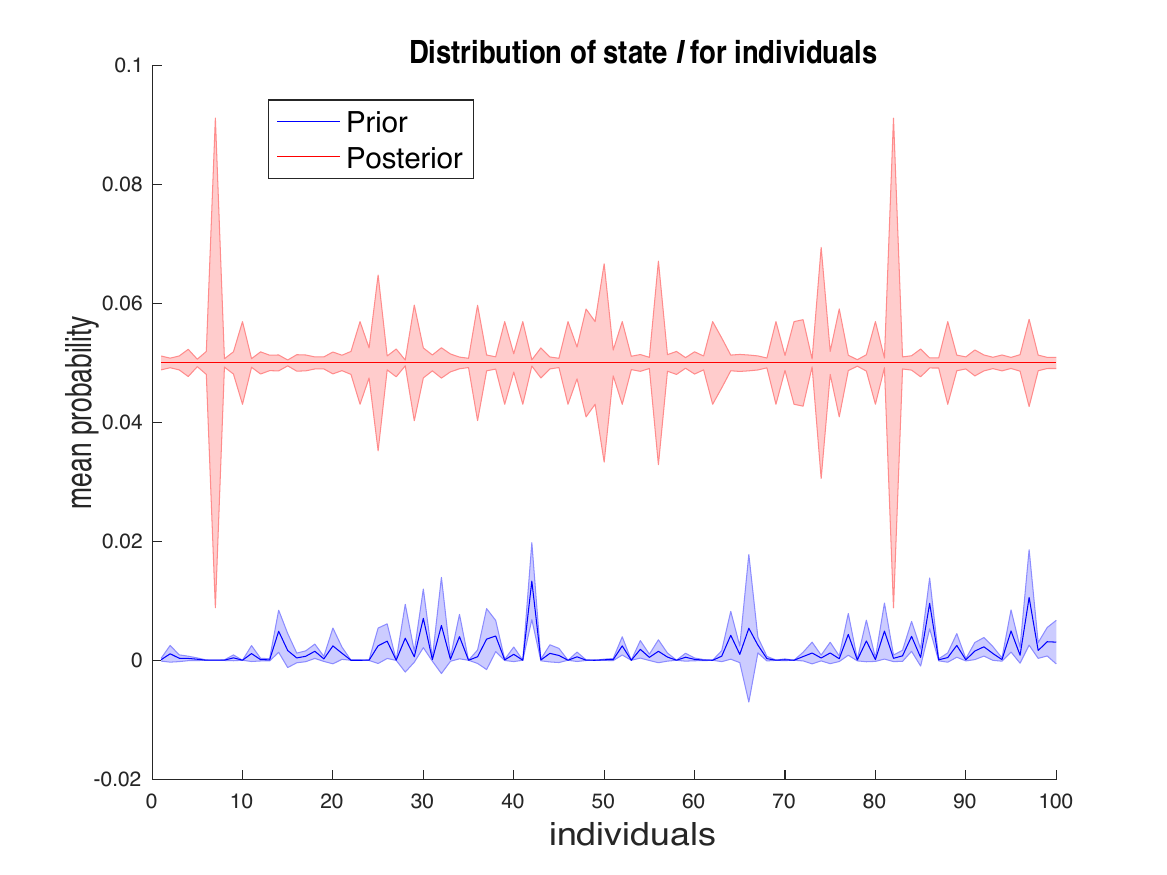}

\caption{PCE and posterior distributions: (Left) Prior and Posterior distributions [mean with standard deviation error band] of the infection state over time steps $t$. (Right) Prior and Posterior distributions of the infection state for 100 individuals. } 
\label{fig:PCE}
\end{figure*}

The last plot depicts the state $\{S,E,I,R\}$ for the 10 individuals  over 20 time instances. We note that the model accounts for both spread of the virus, as well as how the infected individuals recover (and possibly become susceptible again). The rate of change of the states can be optimized by tuning the different parameters (the diffusion coefficients $\kappa_S,\kappa_E, \kappa_I$  and the reaction coefficients $\alpha, \beta, \gamma, \mu_h, \mu_s$) in the model based on data observations, geographical locations, and time. In our experiments, we chose $\kappa_S=0.1,\kappa_E=0.1, \kappa_I=0.25$, and $\alpha=0.02, \beta=0.05, \gamma=0.01, \mu_h =  \mu_s= 0.05$. The statistical distributions for individuals and over time steps are discussed in the next results (see Prior distributions in Figure~\ref{fig:PCE}).  All simulations were performed on Matlab, and  our code {has been made publicly available at \url{https://github.com/Shashankaubaru/GraphSEIR_aPCE}}.

 \begin{figure*}[tb!]
\centering

\includegraphics[height=0.345\textwidth,trim={1cm 1cm 1cm 1cm}]{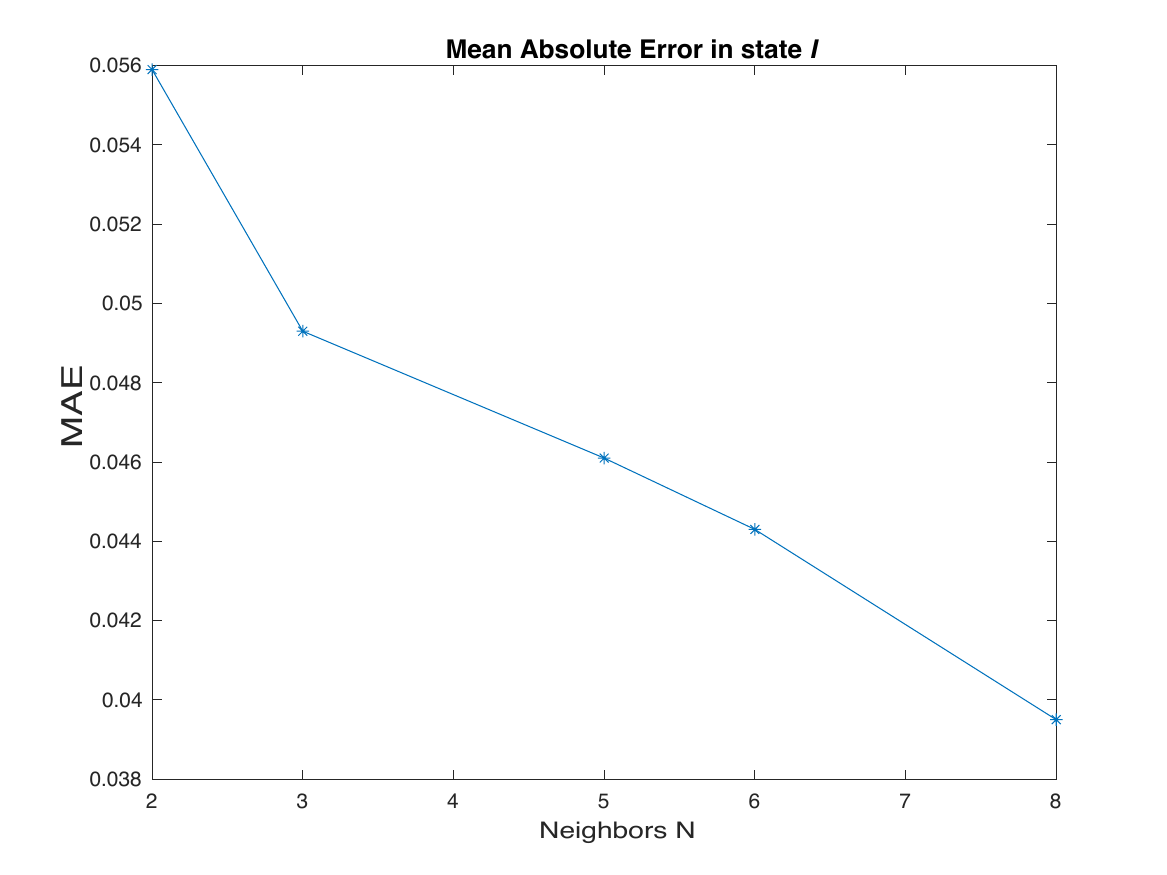}
\includegraphics[height=0.345\textwidth,trim={1cm 1cm 0.5cm 0.5cm}]{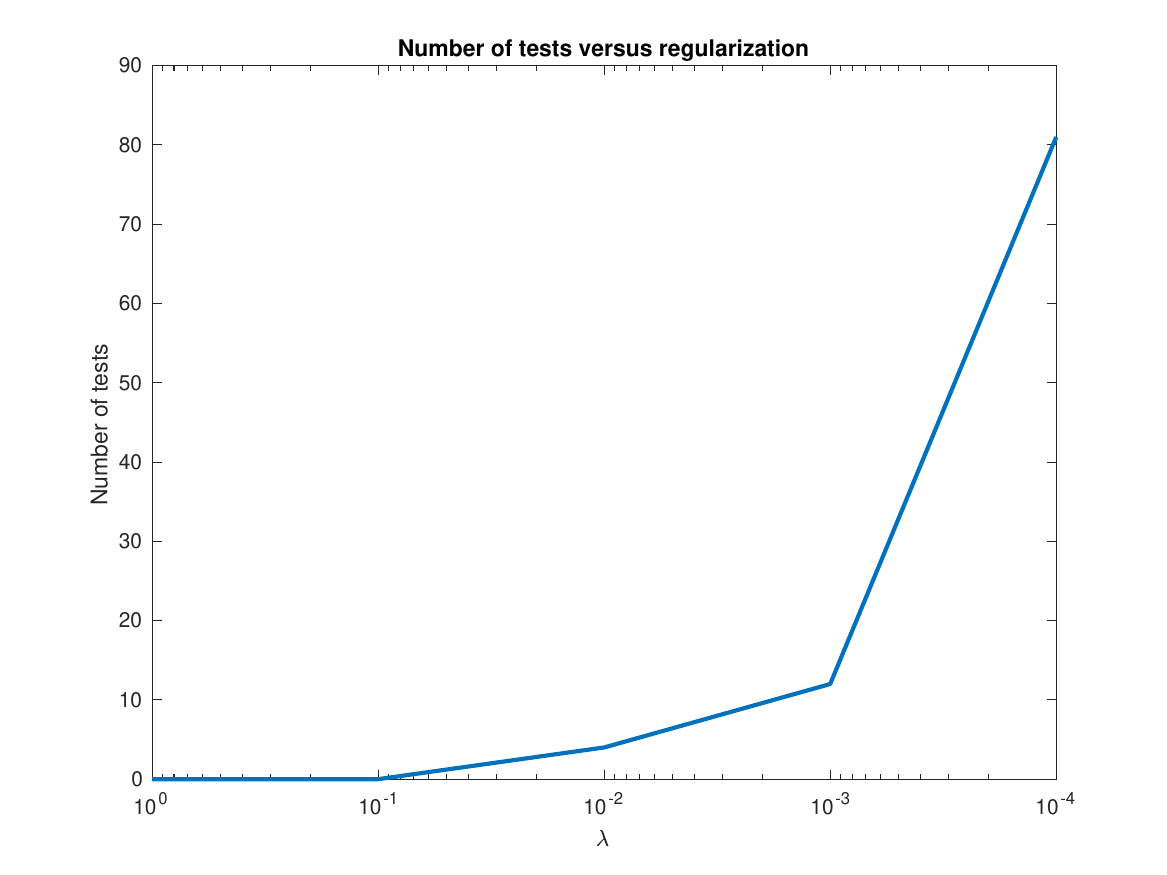}
\caption{PCE and Optimal testing: (Left) The mean absolute error (MAE) between true state $I$ and prediction by PCE as a function of neighbors $N$. { (Right) Number of tests prescribed (cardinality of $\vw_T$) as a function of the regularization parameter $\lambda$.}
}
\label{fig:PCE2}
\vskip 0.1in
\end{figure*}

 \begin{figure*}[tb!]
\centering
\includegraphics[height=0.345\textwidth,trim={1cm 1cm 1cm 1cm}]{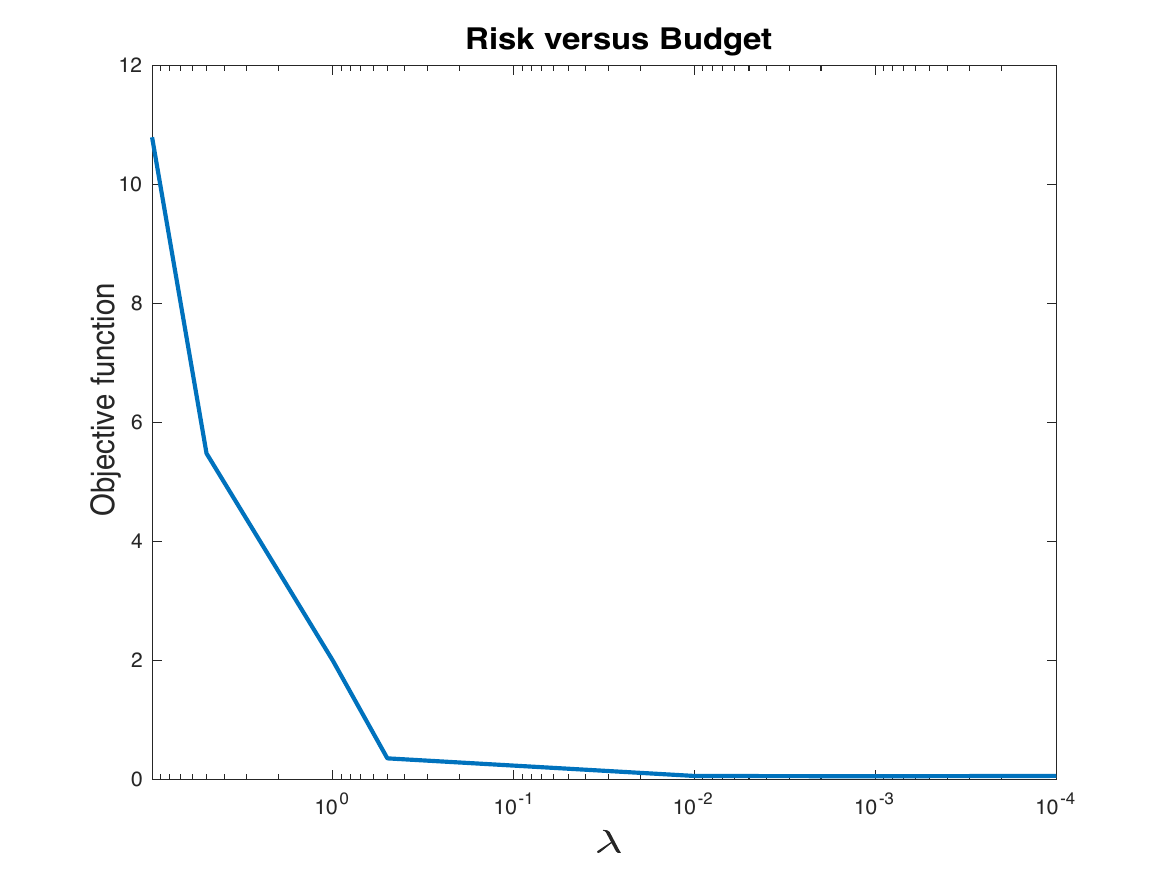}
\includegraphics[height=0.345\textwidth,trim={0.7cm 0.7cm 0.7cm 0.7cm}]{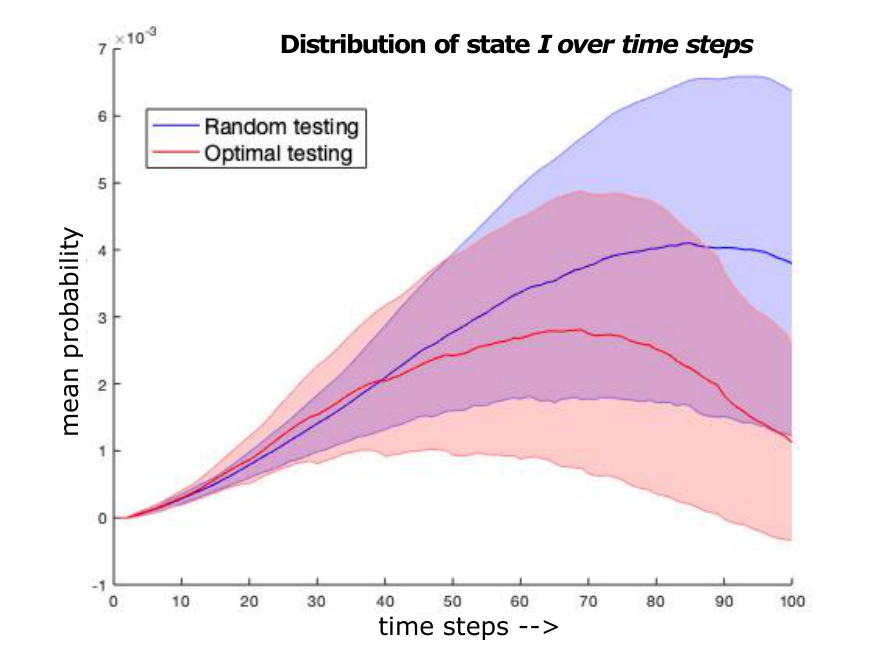}
\caption{PCE and Optimal testing:(Left) Trade-off between the risk (objective function in \eqref{eq:opt}) versus the testing budget (regularization parameter $\lambda$ i.e., no. of tests). (Right) Distribution [mean with standard deviation error band] of state $I$ over time with random and optimal testing.} 
\label{fig:PCE3}
\end{figure*}

\paragraph{PCE and optimal testing:} In the next set of experiments, we study the different aspects of the PCE analysis and uncertainty control. We summarize these results in Figure~\ref{fig:PCE}, \ref{fig:PCE2} and \ref{fig:PCE3}. The first (left) plot in Figure~\ref{fig:PCE}  gives the prior and posterior distributions in the form of the mean with the standard deviation error band of the infection state $I$ over time steps. We considered $n=1000$ individuals to compute the statistics and total time steps $T=100$. The prior distribution is the distribution of the state over time steps as obtained (evolved) from our graphical SEIR model. The posterior distribution is obtained by representing the state using Bayesian aPCE~\cite{article} and computing the response surface using the measurements (uniformly random testing results). We built our PCE simulation using the source code  made available by the authors of~\cite{article}. For PCE, we chose no. of input parameters $N=5$, i.e., we consider $N$ nearest neighbours (based on the edge weights), and the expansion order $d=3$. Hence, the no. of terms (Collocation Points) was $M=56$ ({the} same parameters were used in all experiments).  We observe that the prior distribution is smooth and increasing. This is because the SEIR model does not account for testing. The posterior distribution is random, due to the random testing measurements.
In the second (right) plot, we give the prior and posterior distributions for each individuals obtained from the PCE analysis. We plot the statistics for 100 individuals (we chose fewer nodes for easy visualization) computed over 100 time instances.
Again, the posterior distribution is estimated using the response surface computed using Bayesian aPCE with the above parameters. We observe that the state of certain individuals {has} high variance (high uncertainty). 

In Figure~\ref{fig:PCE2},  the left plot gives us the mean absolute error (MAE) in the prediction of state $I$  by aPCE as a function of the number of neighbors $N$ used to build the expansion. The error is computed as the mean absolute difference of the actual state $I$ as obtained by the SEIR model (considers the whole Laplacian and the test measurements), and the prediction we obtain by aPCE.  For PCE, we assume each state only depends on few neighboring nodes (omitting other nodes and edges), since considering more variables is computationally non-viable. 
We note that, the error reduces as we increase $N$. Increasing $N$ makes the graph more fine-grained, but also increases the complexity of the PCE model. In most situations, the complete contact tracing information/graph will be unavailable, and this result illustrates how our method performs with varying amount of information about the contact network. Similarly, to assess contact tracing information incompleteness, we can drop certain edges at random, depending on the participation rate, when we conduct the PCE analysis. 

In the right plot, {we give the cardinality of the optimal testing prescription vector  $\vw_T$ obtained, i.e, the number of tests prescribed, as a function of the regularization parameter $\lambda$.
The Matlab CVX package~\cite{grant2009cvx} was used to solve the optimization problem in eqn~\eqref{eq:opt} with functions as described before. }
We first observe that, as we decrease $\lambda$, the cardinality of $\vw_T$, i.e., the no. of prescribed tests increases. 
We can choose an optimal $\lambda$ value based on the available budget. Moreover,  we observed that the method prescribes testing for individuals with high uncertainty (individuals with high variance in the right plot of Fig.~\ref{fig:PCE}).
These results show that we can quantify the uncertainty in our model and prescribe appropriate testing.

In figure~\ref{fig:PCE3}, the left plot in the figure presents the \emph{risk to budget trade-off} by plotting the final value of objective function in~\eqref{eq:opt}  we obtained for the optimal $\vw_t$ for different values of the regularization parameter $\lambda$. We again chose $n=1000$, $T= 100$, and other parameters as before. Decreasing  $\lambda$ increases the no. of prescribed tests, and in turn the testing budget required. The plot shows that increasing the no. of tests reduces the risk initially and after a point this reduction is minimal.  The trade-off plot helps us to choose an optimal $\lambda$ (lowest testing budget) for an acceptable risk tolerance. 
The right plot in the Figure~\ref{fig:PCE3}, presents the
distribution of the infection state $I$ over time steps $t$ when testing was conducted randomly (in blue) and when optimal testing was prescribed at regular intervals (in red). We considered $T=100$ time steps, and in the first case, we performed random testing at each time instance. In the second case, we  ran the PCE analysis after every 10 time instances (use previous 10 random measurements to construct the PCE) and used the optimal testing prescription in the next instance. We observe that in the second case, the mean infection starts reducing sooner than the random testing. These results suggest that indeed prescribing optimal testing can help control uncertainty and mitigate disease transmission.

\section*{Conclusions}

 In this study, we introduced a probabilistic SEIR model for disease transmission. The model represents individual-level contact tracing information via dynamic graphs, where each individual represent{s} a node and interaction is described by edges.
 The $S$, $E$, $I$, $R$ compartments are treated as probabilistic entities as to capture uncertainty associated with the stochastic process of disease propagation, sparse  testing, and model inadequacies. 
 As illustrated by numerical simulations, this model can serve healthcare professionals in issuance of early warnings to individuals who are likely exposed or infected by  the virus. Furthermore, the model identifies those individuals who are likely to be asymptomatic. We then proposed the use of arbitrary Polynomial Chaos Expansion (aPCE) to quantify uncertainties in the model, while maintaining computational scalability. By estimating the expected risk as well as  minimizing uncertainty we prescribe optimal testing for  individuals under limited testing and tracing resources. The framework offers a decision tool for balancing between immediate disease spread threat intervention and informed assessment of the pandemic state. Lastly, the framework provides means for policy makers as to estimate the required testing budget for a given acceptable risk tolerance and can be easily adapted  to optimize  vaccine distribution.

\small

\end{document}